\documentclass[twocolumn,aps,prl,amsfonts,amssymb,amsmath,superscriptaddress,groupedaddress]{revtex4}
\usepackage{graphicx}
\usepackage{amsmath}
\usepackage{amsfonts}
\usepackage{amssymb}
\usepackage{graphicx}

\begin{document}

\title{Graphene spintronics: the role of ferromagnetic electrodes}


\author{Jesse Maassen}
\email{maassenj@physics.mcgill.ca}
\affiliation{Centre for the Physics of Materials and Department of Physics, McGill University, Montreal, QC, Canada, H3A 2T8}

\author{Wei Ji}
\email{wji@ruc.edu.cn}
\affiliation{Department of Physics, Renmin University of China, Beijing 100872, China}
\affiliation{Centre for the Physics of Materials and Department of Physics, McGill University, Montreal, QC, Canada, H3A 2T8}

\author{Hong Guo}
\affiliation{Centre for the Physics of Materials and Department of Physics, McGill University, Montreal, QC, Canada, H3A 2T8}

\begin{abstract}
We report a first principles study of spin-transport under finite bias through a
graphene--ferromagnet (FM) interface,
where FM$=$Co(111), Ni(111). The use of Co and Ni electrodes achieves
spin efficiencies reaching $80\%$ and $60\%$, respectively. This large
spin filtering results from the materials specific interaction between
graphene and the FM which destroys the linear dispersion relation
of the graphene bands and leads to an opening of 
spin-dependent energy gaps of $\approx0.4$-$0.5\,{\rm eV}$
at the K points. The minority spin band gap resides higher in energy
than the majority spin band gap located near $\rm E_F$, a feature that
results in large minority spin dominated currents.
\end{abstract}

\maketitle

The field of spintronics, or magneto-electronics, utilizes
the spin degree of freedom of electrons and their inherent
magnetic moment to influence or control the properties of
a circuit. Within this field much effort has
focused on developing interfaces, commonly
a non-magnetic metal or insulator in contact with a
ferromagnet (FM), that exhibits a large spin-polarized
interface resistance\cite{zutic1}. Ideally,
such a spin filter would allow electrons of only
a single spin component to conduct.

Graphene, a 2D lattice of C atoms, is a gapless material with
linear dispersion electronic bands joining
at the Fermi level ($\rm  E_F$) in conical (Dirac) points
located at the K points in the Brillouin zone (BZ) \cite{neto1}.
It has received much attention due to it's exceptional properties\cite{geim1},
including zero effective mass carriers with extremely large mobilities,
and is poised to play a role in the future of nanotechnology.
Among other qualities,
graphene has weak spin-orbit interaction due
to the low atomic number of C resulting
in long spin-flip scattering lengths. Hence, graphene is
a promising material for applications in spintronic devices,
where one can exploit graphene's unique electronic properties within the
context of magneto-electronics.

Generating and injecting a spin-polarized current into graphene is of vital importance 
to the development of graphene-based spintronics. Graphene nanoribbons, unlike
pure graphene, are theoretically predicted to possess a local magnetic moment
at the zigzag edges\cite{sgl-nature}, but a major limitation arises in the difficulty of
reliably fabricating well defined low-disorder edges. Thus, efficient spin injection into 
graphene is required for the realization of a prototypical spintronic device.
A previous first principles study showed extremely large spin filtering efficiencies for 
FM$|$graphene(Gr)$|$FM junctions (FM$=$Co, Ni)\cite{karpan1}, in which the 
current flow was oriented perpendicular to the graphene. In this way, the 
spin-polarized current is primarily dominated by the inter-layer coupling, 
namely van der Waals interactions, between graphene sheets, rather than 
the characteristic graphene states.  
Moreover, the
current-in-plane geometry, with transport occurring parallel to
the graphene, is the most common experimental and theoretical device architecture \cite{exp,semiemp,abinitio,maassen}.
In such systems the source and drain electrodes are comprised of FM-covered graphene 
(because metallic contacts are deposited on top of graphene) which, 
depending on the nature of the chemical bonding,
can hybridize and result in a complex electronic structure.
It was previously shown that graphene placed in contact with
Co or Ni strongly hybridizes leading to significant
modifications of the graphene bands \cite{karpan1}.
Demonstrated by a similar system\cite{ningz}, 
it can be inferred that the interaction between graphene and a FM is 
very sensitive to the particular atomic configuration
at the interface of the two materials. Hence, it is crucial
to properly characterize the detailed atomic structure,
in order to obtain the correct electronic states at the interface and
accurately analyze the spin-polarized transport properties of the device. 
Given the nature of this problem, one must employ atomistic {\em ab initio}
modeling for an accurate treatment of the chemical interaction at the
contact.

\begin{figure}
\includegraphics[width=8.25cm]{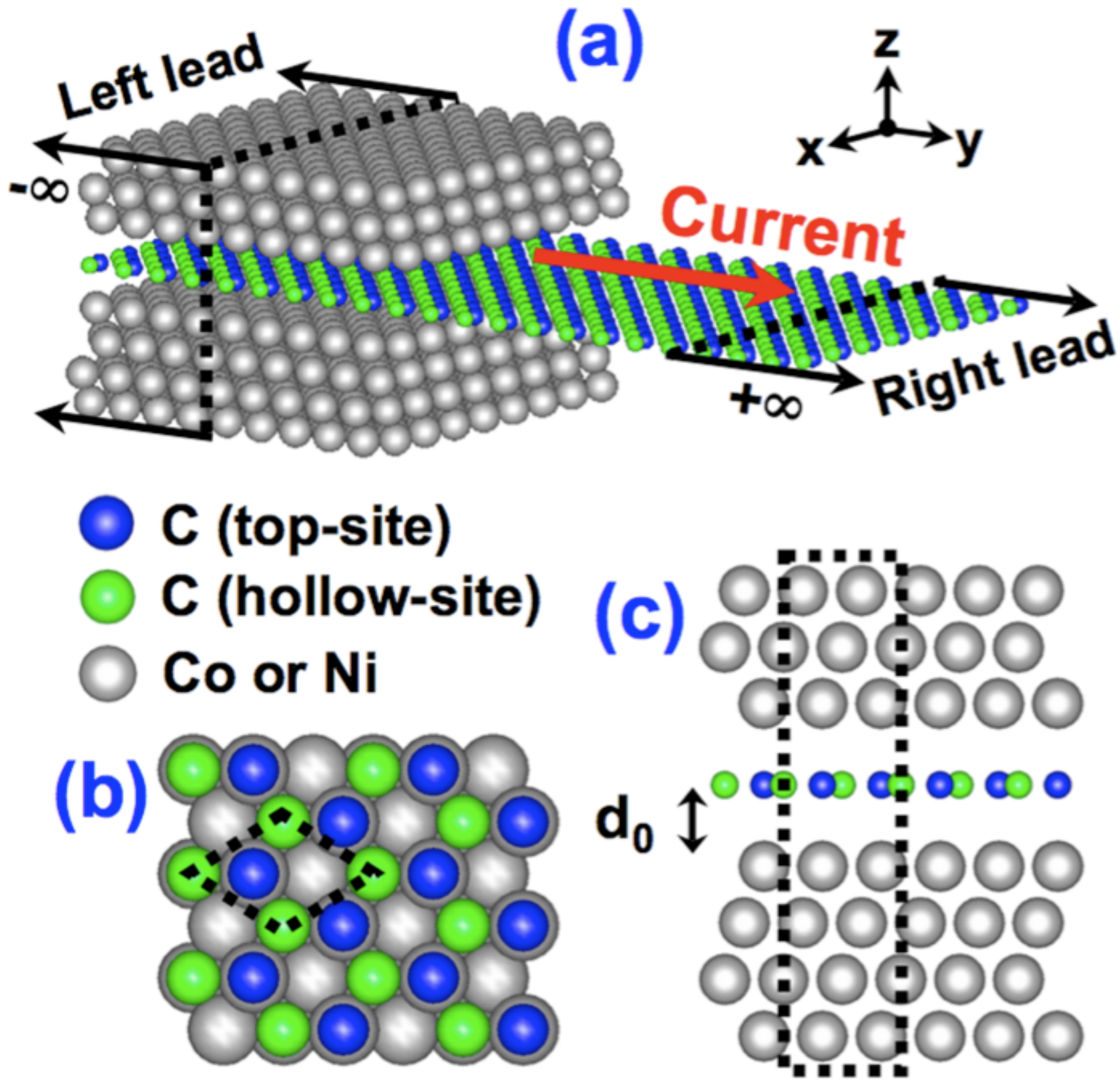}
\caption{(a) Diagram of the Gr$|$Co(111) and Gr$|$Ni(111) interface. The dotted lines
indicate that the left and right leads, extending to $\pm \infty$, consist of
FM-sandwiched graphene (left lead) and pure graphene (right lead).
The top-site and hollow-site C atoms forming the graphene are shown
in blue and green respectively.
Periodic boundary
conditions are assumed in the plane perpendicular to current, i.e.,
the system extends infinitely along the x- and z-directions.
(b) View of the left lead in the x-y plane showing the graphene sitting on the FM.
The top-site C is directly above the FM atom, while the hollow-site C is located at the hollow site.
(c) View of the left lead in the y-z plane indicating the optimized graphene-FM distance $d_0$.
The dotted lines shown in (b) and (c) delimit the supercell box used for the electronic structure calculations.} \label{fig1}
\end{figure}

A nano-structure was constructed to model the interface between a 
source or drain electrode (i.e., the FM-covered graphene)
and a pure graphene channel, as shown in \ref{fig1}. This interface will be
hereafter referred to as Gr$|$FM.
This particular choice in system
geometry was motivated by calculating interface properties that are
independent of the device length. Thus, whether considering a very long (diffusive transport) or
very short (ballistic transport) graphene channel, we expect the results presented
in this work to remain valid at the interface.
We consider both Co(111) and Ni(111) as the FM in
contact with graphene, forming our electrodes. 
For graphene-based spintronics, Co(111) and Ni(111)
are excellent candidates for FM contacts since
their in-plane lattice constants nearly perfectly match that of
graphene\cite{karpan1},
with experimental mismatch values of 1.8\% (Co) and 1.3\% (Ni).

In this Letter, first principles density functional theory (DFT) total energy and non-equilibrium transport calculations were carried out to study the spin-polarized electronic structure and spin-dependent transport properties of Gr$|$FM interfaces under {\em finite bias}. In particular, the atomic structure of the FM-covered graphene was fully relaxed by DFT total energy calculations.  Given the optimized atomic coordinates, the spin-dependent band structure was analyzed. It was found that the strong hybridization between graphene and the FM destroys the linear dispersion relation of the graphene bands and opens spin-dependent band gaps at the K point (similar to previous work \cite{karpan1}). This FM-induced band gap opening of the graphene states results in spin-polarized currents that are
minority (MIN) spin dominated due to the majority (MAJ) spin band gap residing near $\rm E_F$. This electronic feature leads to spin filtering efficiencies reaching above 80\% and 60\% for Gr$|$Co and Gr$|$Ni interfaces, respectively. This work provides a unique analysis of the spin-polarized transport properties of a Gr$|$FM interface while considering the materials specific interactions in a {\em non-equilibrium} setting.

The structural relaxations and band structure calculations were performed using DFT, the local density approximation\cite{lda} for exchange-correlation potentials,
the projector augmented wave method, and a plane wave
basis with a cutoff energy of 400 eV as implemented in the Vienna {\it ab-initio} simulation
package\cite{vasp}. \ref{fig1}(a) shows a diagram of the interface, where the left lead
consists of FM-sandwiched graphene and the right lead is pure graphene. Transport occurs in the
y-direction (as indicated by the red arrow) and periodic boundary
conditions are assumed in the plane perpendicular to current, i.e.,
the system extends infinitely along the x- and z-directions. The most stable (minimal-energy)
configuration for graphene on a Co(111) or Ni(111) substrate was used \cite{karpan1} and corresponds
to the top-site C located directly above the FM atom and the hollow-site C sitting
at the hollow site (see \ref{fig1}(b) and \ref{fig1}(c)). Seven FM layers are used to separate the graphene sheets in adjacent supercells along the z-direction. The atomic structure
of the graphene-FM contact was obtained by fixing the in-plane lattice constant
to graphene's value of 2.46~\AA~and relaxing the atoms in the supercell until the net 
forces acting on the atoms were below 0.01 eV/\AA. The supercell box height, in the 
direction perpendicular to graphene,
was varied after each relaxation in order to find the optimal height through total energy minimization. 
A $k$-mesh of 21$\times$21$\times$3 was adopted to sample the BZ for 
structural relaxations and total energy calculations. These parameters 
provide the optimal graphene-FM distance $d_0$ equal to  
2.17~\AA~and 2.13~\AA~for Gr$|$Co(111)
and Gr$|$Ni(111) respectively.

\begin{figure}
\includegraphics[width=8.25cm]{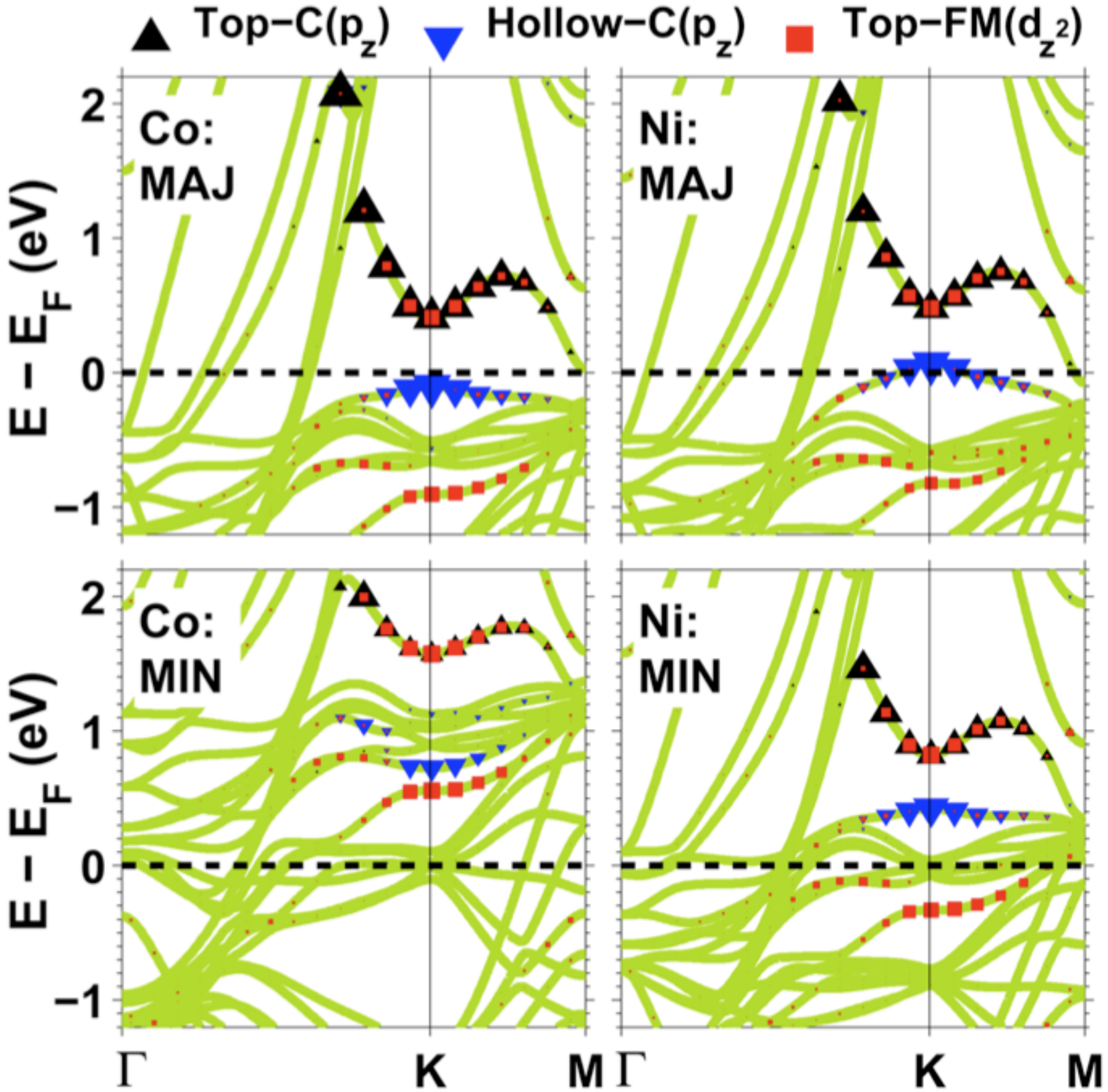}
\caption{Electronic structure of Co(111)- and Ni(111)-sandwiched graphene (left lead of system in \ref{fig1}(a)).
The left and right panels show the majority spin and minority spin bands for Co and Ni respectively.
The Fermi level is indicated by the horizontal dashed line. The black up-triangles (blue down-triangles) show the $\rm p_z$ character of the top-site (hollow-site) C, while the red diamonds present the $\rm d_{z^2}$ character
of the Co or Ni atoms located directly above and below the top-site C.} \label{fig2}
\end{figure}

The spin-dependent band structure of the left lead is shown in \ref{fig2}.
The green lines correspond to the bands of the
hybrid graphene-FM system. To locate the states originating from graphene,
the weight of the C($\rm p_z$) orbitals
of all bands is superimposed (black up-triangle: top-site C, blue down-triangle: hollow-site C).
There are three distinct features found from the figure:
(i) At the K point, the graphene bands no longer show a linear
dispersion relation and now exhibit a band gap opening of $\approx$ 0.4-0.5~eV, 
similar to what was shown for a single FM surface in contact with graphene \cite{karpan1}.
The majority spin band gap resides at a lower energy than 
the minority spin band gap, and they do not overlap. This
indicates that at specific energies, one finds electrons
of only a single spin type in the graphene.
(ii) Superimposing the $d_{z^2}$ character of the FM atoms located
in the first layer above and below the graphene (plotted as red squares)
shows that the FM interacts only via the top-site C. This can be seen
from the fact that the red squares only overlap with the black up-triangles.
Note that only the $d_{z^{2}}$
orbitals showed any significant mixing with the graphene, in accordance with
previous work \cite{karpan1}.
(iii) It is also clear that the graphene conduction band (CB) and valence band (VB)
result separately from the top-site C and the hollow-site C,
respectively. This is in contrast to pure graphene, where
both C atoms in the primitive cell equally contribute to the
CB and VB. The C-site dependence on the graphene CB and VB
can be understood from the graphene-FM interaction (mentioned in (ii))
which breaks the sub-lattice symmetry between the top-site C and the
hollow-site C.

\begin{figure}
\includegraphics[width=8.25cm]{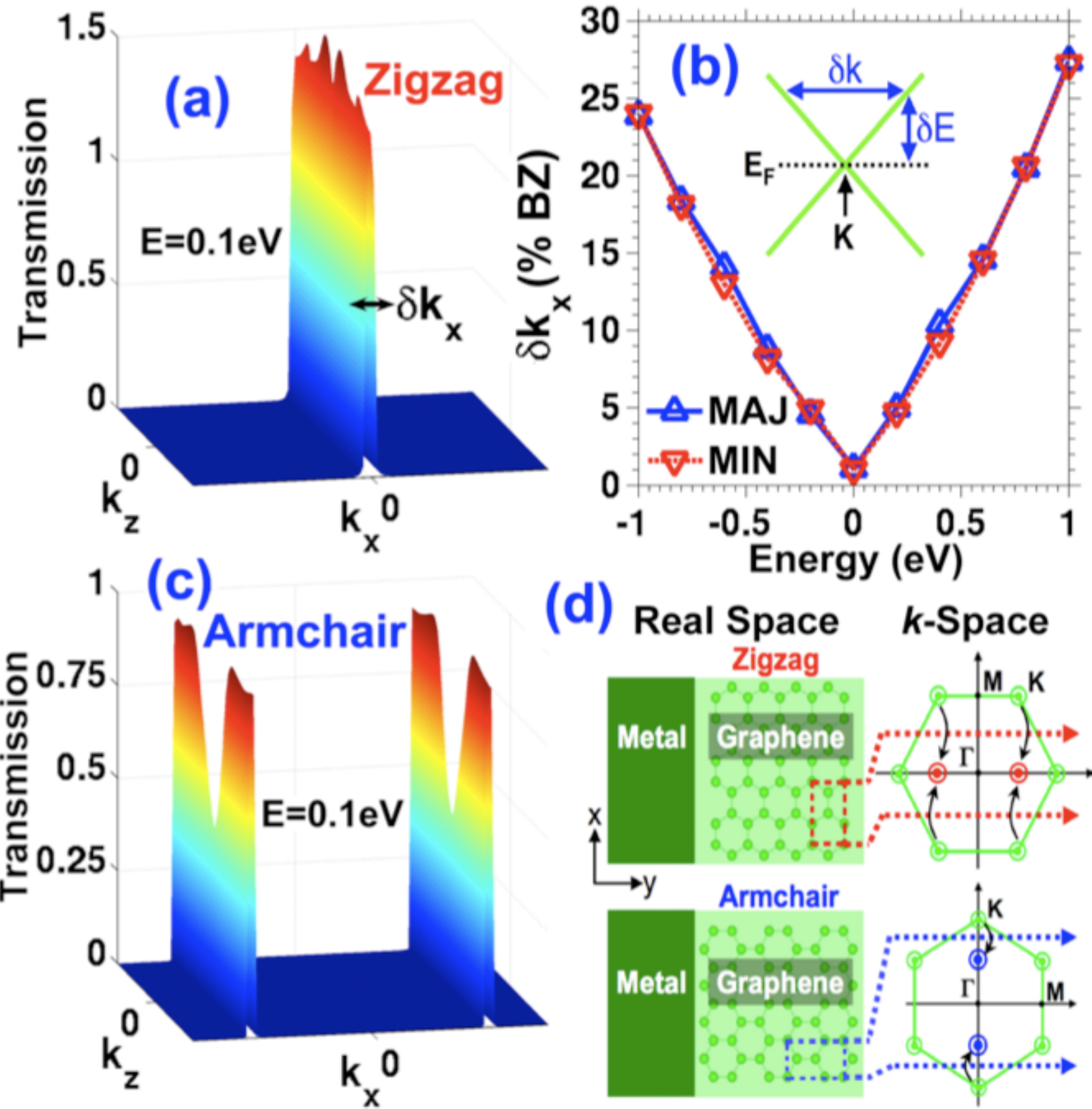}
\caption{(a) $T_{MIN}$ versus $k_{\perp}$ at $E=0.1\,{\rm eV}$ for
Gr(zigzag)$|$Ni. (b) Spin-polarized $\delta k_x$ versus $E$
for the same system as in (a). $\delta k_x$ is defined as the width at half-max of $T$.
The inset illustrates the linear $E$-$k$ dispersion bands of graphene.
(c) $T_{MIN}$ versus $k_{\perp}$
at $E=0.1\,{\rm eV}$ for Gr(armchair)$|$Co.
(d) Left panels: Top view in real space of Gr(zigzag)$|$FM and
Gr(armchair)$|$FM. Right panels: Primitive BZ of
pure graphene(zigzag) (top) and graphene(armchair) (bottom). The dashed arrows indicate
the transport direction as well as delimit the zone edges along $k_x$ for our calculations,
due to the use of a non-primitive supercell (the supercell width along x is shown in the
left insets).
The black curved arrows illustrate the folding of the K points.}
\label{fig3}
\end{figure}

Given the fully relaxed atomic structures and the well understood electronic states of the left lead, we employed our state-of-the-art {\em ab initio} transport package, named {\sc MatDCal} \cite{matdcal}, to compute the spin-dependent transport properties of the whole junction. {\sc MatDCal}
uses non-equilibrium Green's functions
(NEGF) combined with DFT for open systems in a two-probe geometry under
finite bias, where the leads extend to $\pm \infty$. An optimized double-$\zeta$ polarized
atomic orbital set was built for each atomic species. The local density approximation \cite{lda}
and norm-conserving non-local pseudopotentials \cite{pseudo} were used
and $k$-point convergence tests, including the high symmetry points ${\rm \Gamma}$,
K and M, were performed for all calculations.

The spin-dependent transmission coefficient of the minority states ($T_{MIN}$)
in the 2D BZ (in the plane of $k_{x}$ and $k_{z}$ which are $\perp$ to the transport direction, i.e., y direction) for Gr(zigzag)$|$Ni is plotted in \ref{fig3}(a).
The notation graphene(zigzag) or graphene(armchair) indicates the graphene is oriented such
that transport occurs along the zigzag or armchair direction, respectively.
$T_{MIN}$ shows a sharp peak at $k_x=0$, with near-zero values away from this point.
To quantify this unusual dependence on $k$, 
we measure the width ($\delta k_x$)
of the $T_{MIN}$ peak. $\delta k_x$ is defined as the half-max width of $T$,
as depicted in \ref{fig3}(a). \ref{fig3}(b) presents the calculated spin-polarized
$\delta k_x$ as a function of energy ($E$) for Gr(zigzag)$|$Ni, which clearly shows a nearly linear behavior.
The spread in $\delta k_x$ can be traced back to the conical states of
pure graphene, which also exhibit a linear $E$-$k$ relationship as shown in the inset of \ref{fig3}(b). For a Gr$|$FM interface, the pure graphene forming the 
right lead of our system only has
electronic states at the K points (for $E\approx {\rm E_F}$). Hence, all the incoming
carriers originating from the FM-covered graphene (i.e., the source or drain electrode)
are blocked by the pure graphene except those with states at the K points.
\ref{fig3}(c) presents $T_{MIN}$ versus $k_{\perp}$ for Gr(armchair)$|$Co.
The armchair-oriented graphene also shows sharp peaks along $k_x$, but are shifted
in comparison to the zigzag-oriented graphene to $k_x=\pm2/3$. The positions of the peaks
along $k_x$ can be explained, for both graphene orientations, by considering the
effect of band folding on the graphene states. In \ref{fig3}(d), we show the primitive BZ
of graphene, where the dashed arrows indicate the transport direction in addition to
delimiting the $k_x$ zone edges due to our non-primitive supercell.
The K points located outside the $k_x$ zone edges will be folded inward, as
illustrated with the black curved arrows. For graphene(zigzag), the conducting K points all
appear at $k_x=0$ leading to a single peak in $T_{MIN}$, while the K points for
graphene(armchair) are positioned at $k_x=\pm 2/3$, as seen in \ref{fig3}(a) and \ref{fig3}(c).

\begin{figure}
\includegraphics[width=8.25cm]{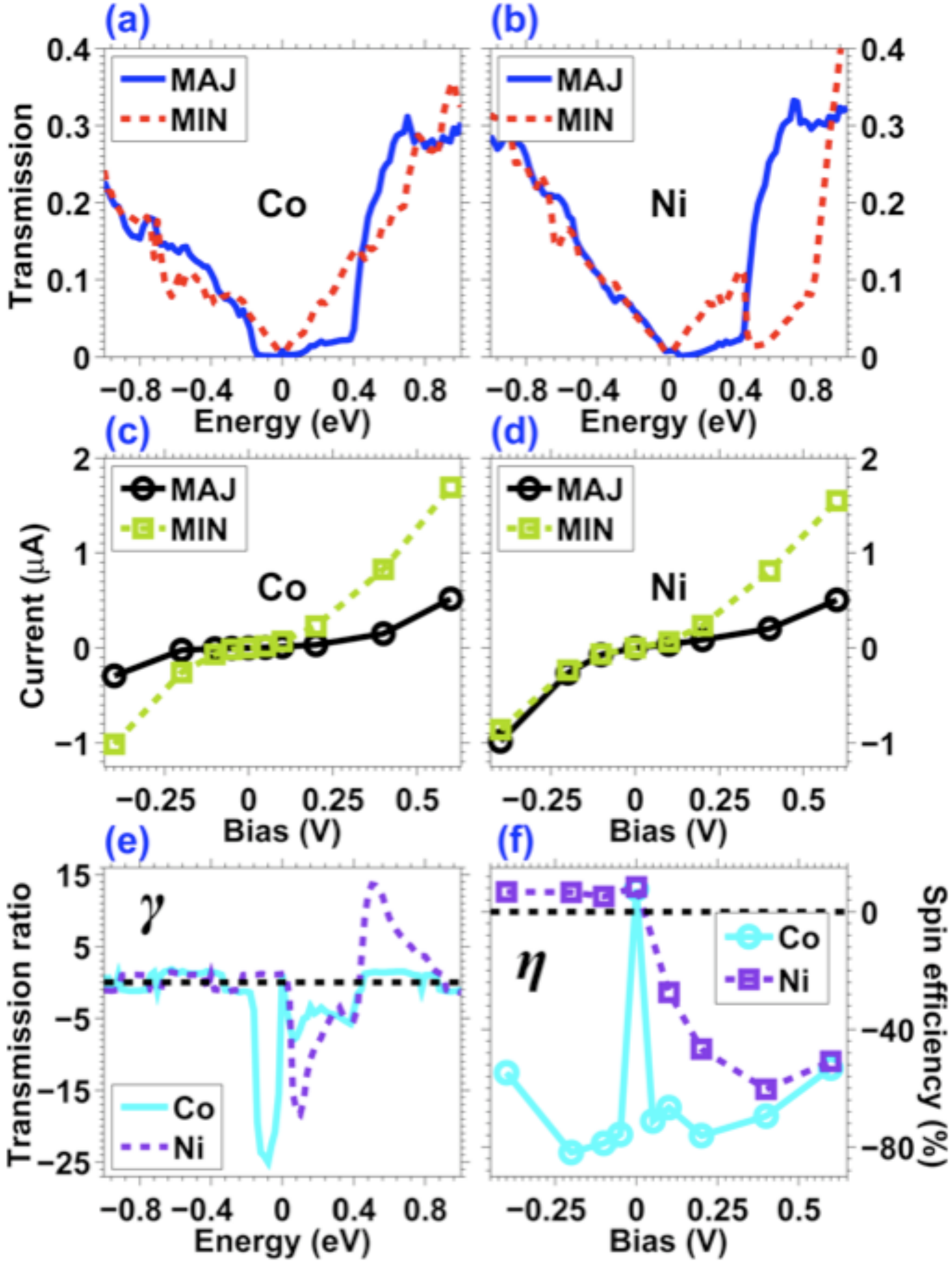}
\caption{(a)-(b): Spin-polarized $T$ versus $E$
at $V=0$ for Gr$|$Co (a) and Gr$|$Ni (b) (majority spin: solid blue
line, minority spin: dashed red line). The Fermi level is set to $E = 0$.
(c)-(d): Spin-polarized $I$ versus $V$ for Gr$|$Co (c) and Gr$|$Ni (d)
(majority spin: black circles, minority spin: green squares).
(e) Transmission ratio $\gamma \equiv T_{\sigma}/T_{\bar{\sigma}}$
(where $\sigma$ is the spin component with the largest $T$,
and $\bar{\sigma}$ is the opposite spin) versus $E$
for Gr$|$Co (cyan full line) and Gr$|$Ni (purple dashed line).
Note that $T_{\sigma}/T_{\bar{\sigma}}$
is defined positive (negative) when $T_{MAJ}>T_{MIN}$ ($T_{MIN}>T_{MAJ}$).
(f) Spin efficiency $\eta \equiv |I_{\sigma}-I_{\bar{\sigma}}|/|I_{\sigma}+I_{\bar{\sigma}}|$ versus $V$
for Gr$|$Co (cyan circles) and Gr$|$Ni (purple squares).
Spin efficiency is defined positive (negative) when
$|I_{MAJ}|>|I_{MIN}|$ ($|I_{MIN}>I_{MAJ}|$).}
\label{fig4}
\end{figure}

The $k$-averaged spin-dependent transmission coefficients ($T$) versus energy $E$ at equilibrium (zero bias) for Gr(zigzag)$|$Co(Ni) are presented in \ref{fig4}(a) and \ref{fig4}(b), which were calculated using $T_{\sigma}(E)=1/A_{BZ} \int_{BZ} T_{\sigma}(E,k_{\perp})\,dk_{\perp}$, where $A_{BZ}$ is the area of the BZ.
For $E$ smaller than $\approx {\rm E_F}$
($\rm E_F$ is set to zero), it is found that both $T_{MAJ}$ and $T_{MIN}$
vary roughly linearly, similar to pure graphene. However,
the $T$ values for Gr$|$FM are roughly 50\% smaller than those of pure graphene,
a result of the band gap opening which reduces the band velocity.
When considering $E>{\rm E_F}$, one noticeable feature for both Gr$|$Co
and Gr$|$Ni is the small $T_{MAJ}$ value
between $E\approx0$-$0.4\,{\rm eV}$. This energy range corresponds to the majority
spin band gap of the graphene states (see \ref{fig2}), thus explaining the decrease in $T_{MAJ}$.
A similar effect occurs for Gr$|$Ni but with $T_{MIN}$ (instead of $T_{MAJ}$)
between $0.43$-$0.80\,{\rm eV}$,
an energy range which overlaps with the minority spin band gap.
\ref{fig4}(e) presents the spin transmission ratio ($\gamma$) defined as $T_{\sigma}/T_{\bar{\sigma}}$,
where $\sigma$ is the spin component with the larger $T$ value and $\bar{\sigma}$ is the opposite spin.
$\gamma$ is defined positive (negative) when $\sigma$ is the majority (minority) spin. 
Gr$|$Co and Gr$|$Ni
both yield large dips near $-0.1\,{\rm eV}$ and $0.1\,{\rm eV}$ with $\gamma$ approaching $18$ and $25$ respectively,
each coinciding with the smallest $T_{MAJ}$ value. Interestingly, for Gr$|$Ni, $\gamma$ shifts from minority
spin dominated at $0.1\,{\rm eV}$ to majority spin dominated at $0.5\,{\rm eV}$. This crossover behavior
is attributed to the end of the majority spin band gap and the beginning
of the minority spin band gap both located near $0.4\,{\rm eV}$.

The non-equilibrium calculations reveal the spin-polarized current ($I_{\sigma}$)-voltage ($V$)
characteristics of the Gr$|$Co and Gr$|$Ni interfaces, as plotted in \ref{fig4}(c) and (d). $I_{\sigma}$ is obtained from
\begin{equation}
I_{\sigma}=\frac{e}{h} \int_{-\infty}^{\infty} T_{\sigma}(E)[f_L(E,\mu_L)-f_R(E,\mu_R)]\,dE,
\end{equation}
where $e$ is the electron charge, $h$ is Planck's constant and
$f(E,\mu)$ is the Fermi-Dirac distribution.
An applied bias $V$
varies the left and right chemical potentials as $\mu_L = E_F$
and $\mu_R = E_F + |e|V$, where $V = V_L - V_R$.
For Gr$|$Co, it was found that $|I_{MIN}|>|I_{MAJ}|$ for 
all $V$ values in the bias voltage window of interest.
Whereas in the case of Gr$|$Ni, $|I_{MIN}|>|I_{MAJ}|$ is only observed
for positive $V$ while the opposite result
is found for negative $V$ (although less pronounced). To illustrate this,
we plot the spin efficiency ($\eta$) for the Gr$|$FM interfaces in \ref{fig4}(f).
$\eta$ is calculated using 
$|I_{\sigma}-I_{\bar{\sigma}}|/|I_{\sigma}+I_{\bar{\sigma}}|$,
where $\eta$ is defined positive (negative) when
$|I_{MAJ}|>|I_{MIN}|$ ($|I_{MIN}|>|I_{MAJ}|$). Gr$|$Co and Gr$|$Ni achieve maximal spin
efficiencies above $80\%$ (at $-0.2\,{\rm V}$) and $60\%$ (at $0.4\,{\rm V}$) respectively, 
representing the percentage of net spin-polarized current.
$\eta\rightarrow 0$ for negative $V$ in the case of Gr$|$Ni.
This occurs because the integration window of $E$ for calculating $I_{\sigma}$
(ranging from $({\rm E_F}+|e|V)\rightarrow{\rm E_F}$, valid for $V<0$ at zero temperature)
is located below both majority and minority spin
band gaps (see \ref{fig2}). This results in near-equal currents from both spin types.
For Gr$|$Co, $\eta$ saturates slowly to zero with decreasing $V$,
because when the integration range extends to the bottom of the majority spin band gap,
located near $-0.15\,{\rm eV}$, $T_{MAJ}$ remains small due to the vanishing
density of states at the Dirac point in the pure graphene (which is pinned at the lower boundary of the integration window). Hence, $V$ must decrease beyond
$-0.15\,{\rm V}$ to obtain $\eta\rightarrow0$. The peak at $V=0$
results from $I_{MAJ}$ and $I_{MIN}$ both vanishing in pure graphene.
The results shown in \ref{fig4} have
considered only graphene(zigzag). The transport properties of graphene(armchair) are found to be qualitatively similar to graphene(zigzag).

In summary, our non-equilibrium first principles transport calculations
showed that Gr$|$Co(111) and Gr$|$Ni(111) interfaces exhibit large spin injection
values reaching $80\%$ and $60\%$ respectively. This effect originates from the
graphene-FM hybridization which leads to an opening of the conical Dirac bands
resulting in spin-dependent energy gaps of the
graphene states. Thus, in the ballistic regime, one can (theoretically) exploit
the materials specific bonding between graphene and the FM to achieve very efficient
spin filtering. However, it remains to be shown whether these spin properties are 
robust in the inevitable presence of random interface disorder.


{\em Acknowledgements.}
This work was supported by the FQRNT
of Quebec, NSERC of Canada and CIFAR. Calculations were performed
using the RQCHP supercomputers.




\end{document}